\documentclass[journal,twoside,web]{ieeecolor}
\usepackage{tmi}
\usepackage{cite}
\usepackage{amsmath,amssymb,amsfonts}
\usepackage{algorithm}
\usepackage{algorithmic}
\usepackage{booktabs}
\usepackage{pifont}
\usepackage{graphicx}
\usepackage{textcomp}
\usepackage{CJKutf8}
\usepackage{stfloats}
\usepackage{multirow}
\usepackage{adjustbox}
\def\BibTeX{{\rm B\kern-.05em{\sc i\kern-.025em b}\kern-.08em
    T\kern-.1667em\lower.7ex\hbox{E}\kern-.125emX}}
\begin{document}
\begin{CJK}{UTF8}{gbsn}
\title{2D Sparse Array Design via Reweighted L1 Second Order Cone Programming for 3D  Ultrasound Imaging}
\author{Xi Zhang, Miguel Bernal, and Wei-Ning Lee, \IEEEmembership{Member, IEEE}
\thanks{This paragraph of the first footnote will contain the date on which
you submitted your paper for review. It will also contain support information,
including sponsor and financial support acknowledgment. For example, 
``This work was supported in part by the U.S. Department of Commerce under Grant BS123456.'' }
\thanks{The next few paragraphs should contain the authors' current affiliations,
including current address and e-mail. For example, F. A. Author is with the
National Institute of Standards and Technology, Boulder, CO 80305 USA (e-mail:author@boulder.nist.gov). }
\thanks{S. B. Author, Jr., was with Rice University, Houston, TX 77005 USA.
He is now with the Department of Physics, Colorado State University,
Fort Collins, CO 80523 USA (e-mail: author@lamar.colostate.edu).}
\thanks{T. C. Author is with the Electrical Engineering Department,
University of Colorado, Boulder, CO 80309 USA, on leave from the National
Research Institute for Metals, Tsukuba, Japan (e-mail: author@nrim.go.jp).}}

\maketitle

\begin{abstract}
Two-dimensional (2D) fully-addressed arrays can conveniently realize three-dimensional (3D) ultrasound imaging while fully controlled such arrays usually demands thousands of independent channels, which is costly. Sparse array technique using stochastic optimization methods is one of promising techniques to reduce channel counts while due to the stochastic nature of these methods, the optimized results are usually unstable. In this work, we introduce a sparse array design approach that formulates the synthesis problem of sparse arrays as second-order cone programming (SOCP) and a re-weighted L1 technique is implemented to sequentially optimize the SOCP. Based on this method, an on-grid quasi-flatten side-lobe (Q-Flats) 2D sparse array with side-lobe level (SLL) no more than -21.26 dB and 252 activated elements is designed, which aims to achieve as high contrast performance as possible under the limits of resolution and maximum number of independent channels (i.e., 256). The imaging performance of the Q-Flats array was compared with those of a corresponding dense array (Dense), a Fermat spiral array (Spiral) and a spatially 50$\%$-Tukey tapered spiral array (Spiral-Taper) using Field II simulations in a multi-angle steered diverging wave transmission scheme. It was demonstrated that the Dense achieved the best resolution and contrast and the Spiral-Taper the worst. The Q-Flats showed better resolution (about 3$\%$) but slightly worse contrast than the Spiral.   
All the results indicate the re-weighted L1 SOCP method is a promising and flexible method for seeking trade-offs among resolution, contrast, and number of activated elements.

\end{abstract}


\section{Introduction}
\label{sec:introduction}
\IEEEPARstart{U}{trasound} volumetric imaging inherently provides three-dimensional (3D) information of tissue structures and dynamics and reduces inter-operator and diagnostic variability, thereby preferred over two-dimensional (2D) ultrasound imaging \cite{68466,fenster2001three,1293556}. Temporal evolution of 3D ultrasound imaging is particularly valuable for clinical applications, such as tracking and guiding of surgical instruments during robotic beating-heart intracardiac surgery surgeries \cite{N6880838}, detection of the brain activity \cite{GESNIK2017267,rau20183d}, assessment of mechanical properties of tissues \cite{deffieux2008assessment,deprez20093d,gennisson20154,papadacci20163d,correia20183d}, reconstruction of spatiotemporal  distributions of blood flow and micro-vessels \cite{7185013,provost20153,rossi2020high,demeulenaere2022vivo}, and monitoring of High Intensity Focused Ultrasound (HIFU) treatment \cite{unsgaard2006intra}. 

Volumetric ultrasound imaging can be realized by manual or mechanical translation/rotation of one-dimensional (1D) array probes \cite{5587409, nikolov2002three} or a direct use of 2D array probes. Volumetric ultrasound imaging with motorized 1D arrays can lead to poor spatial resolution in the elevational direction and a low frame rate. Ideally, 2D matrix probes can focus or steer ultrasonic beams in both the elevational and azimuthal directions \cite{84270} and achieve high volume rate by transmitting unfocused beams such as plane waves \cite{4816058,provost20143d} or diverging waves\cite{gennisson20154}. In the case of beam steering, the element pitch in full arrays is required to be less than half of the wavelength to prevent grating lobes from appearing in the desired field of view. This means that a 2D array should contain thousands of elements in order to produce comparable image quality in both azimuth and elevation to the lateral direction of a 1D array. In other words, driving a large dense array needs imaging system with thousands of independent channels, which poses significant challenges on not only fabrications, but system cost,  size of coaxial cables and data transfer rate \cite{7549037}. Besides, current standard ultrasonic imaging systems can only support up to 256 independent channels. Despite four Vantage 256 scanners have been successfully synchronized to drive a 32$\times $32 fully addressed array \cite{petrusca2018fast}, it is of significance to explore channel reduction techniques since it is costly and inconvenient. 

One possible solution is micro-beamforming \cite{5306761, kortbek2013sequential, 6805693}. Unlike conventional beamforming performed at the imaging system backend, micro-beamforming divides the beamforming process into two stages and move the first stage into the probe handle. In the first stage, the fine delays are applied to the received echo signals from the predetermined subarray elements through application-specific integrated circuits (ASIC) inside the probe, which reduces the channel counts. In the second stage, coarse delays are applied to the pre-beamformed signals produced by the first stage in the main system. However, integrating ASIC is expensive and increases the probe fabrication challenges.

Another attractive solution is 2D row-column addressing (RCA) arrays \cite{1293560, 4815314, 5075108, 5895041, 7103534, christiansen20153}. In RCA,  elements along a row or column are treated as 1D linear array through crossed electrodes. Thus, the number of required channels is reduced by a factor of $N/2$ (from $N^{2}$ to 2$N$). During imaging, ultrasound beams are transmitted along either rows or columns, and then received along the orthogonal electrodes. As a result, transmit focusing can only be enabled in one direction, and receive focusing in the orthogonal direction. However, the imaging region of RCA is rectilinear in front of the array, which is not

\begin{figure}[!t]
	\centering
	\includegraphics[width=\columnwidth]{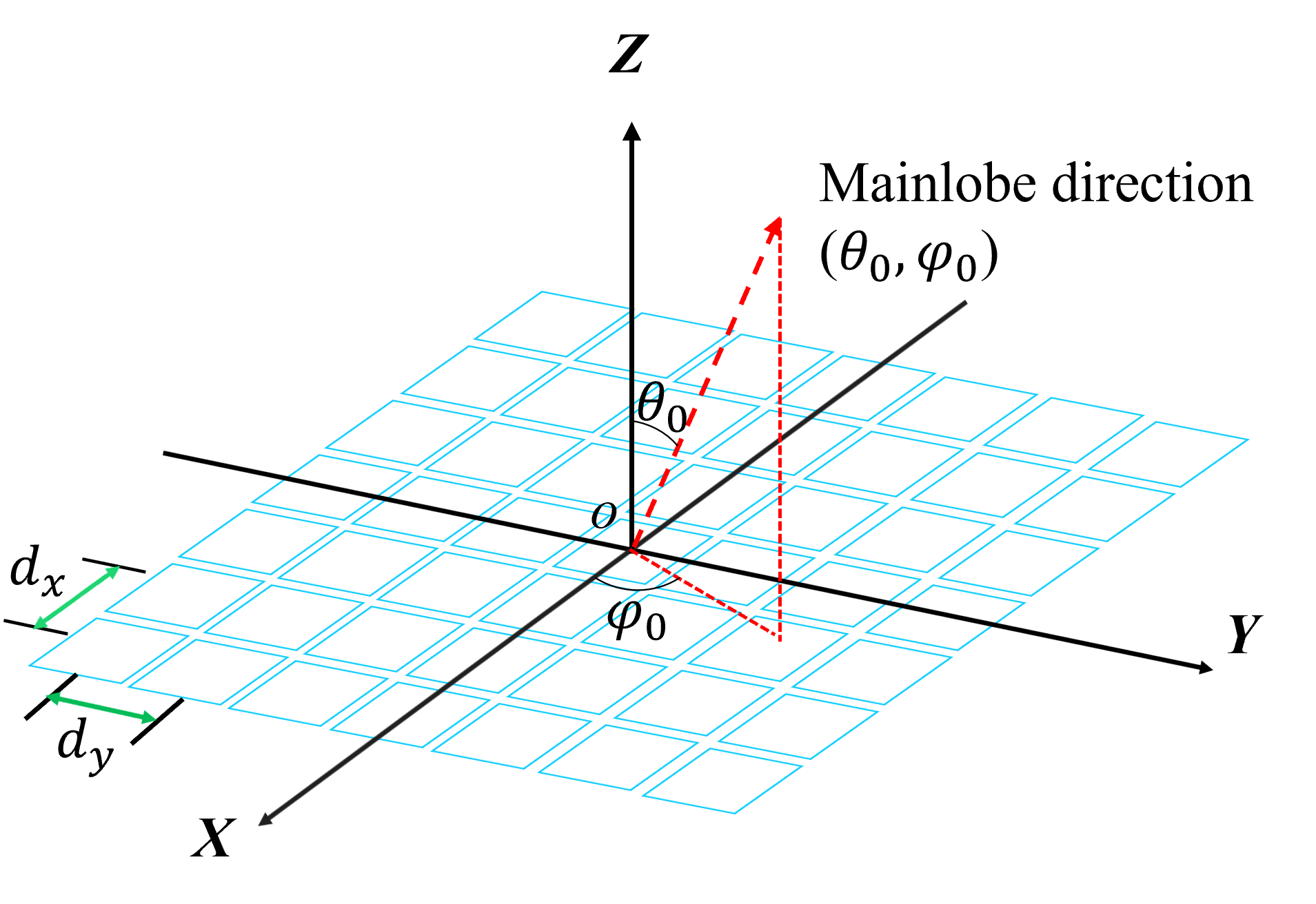}
	\caption{The layout of 2D rectangular dense array. $d_x$, and $d_y$ denote element pitches in azimuth and elevation, respectively.}
	\vspace{-1em}
	\label{2D array.}
\end{figure}
\noindent suitable for volumetric phased array imaging (e.g., heart imaging), unless with a diverging lens \cite{8092155}.

Sparse array \cite{9741756} is a promising technique to channel reduction by directly reducing the number of active elements that is needed to be independently controlled by the imaging system \cite{171469}. The core idea of sparse array is to reduce elements while maintaining the array performance at a certain level in terms of resolution, side-lobe level (SLL) and grating lobe level. Generally, approaches used to design sparse array can be classified into two categories:  $deterministic$ \cite{nikolov2000application, brunke2002broad, lookwood2002optimizing, austeng2002sparse, martinez20102d, ramadas2014application, ramalli2015density, masoumi2023costas} and $non-deterministic$ methods \cite{84270, 299602, 753023, 6529099, 7549037, 7581110}, respectively. Deterministic methods that construct sparse array through explicit regularities or mathematical formulas include regular and radially sparse periodic array \cite{austeng2002sparse}, Vernier array \cite{brunke2002broad}, and Mills cross arrays \cite{818752}. Recently, Spiral \cite{martinez20102d, ramalli2015density} and Costas arrays \cite{masoumi2023costas} have also been introduced and gained much attention. Fermat's spiral array, also known as sunflower array, is defined by sampling the Fermat's spiral equation. By setting the so-called divergence angle as golden angle that is an irrational number, each element will have a different angular position from another one, thus breaking any periodicity in the array. As pointed by Sumanaweera et al.\cite{849228}, spiral array designs can minimize element shadowing and thus contribute to low SLL. Density tapering techniques \cite{vigano2009sunflower, 5067734} which act as apodization contracts the elements towards the aperture center to reduce the SLL at the cost of resolution. An $N\times N$ Costas sparse 2D array \cite{masoumi2023costas} is selected from all the order-$N$ Costas arrays \cite{beard2007costas, h21p42-17}. In a Costas sparse array, any two elements can not be presented in the same row or column and has a distinct displacement vector between them, which makes it promising to simplify the circuit interconnections. 

For the non-deterministic 2D sparse array design methods, randomly deactivated some of the elements in a corresponding dense array is initially attempted \cite{84270}. However, since finding an optimal layout among all possible sparse arrays is a very large combinatorial optimization problem, this random-based method usually cannot obtain the best choice. Based on this, stochastic optimization methods, such as genetic algorithms (GA) \cite{299602}, and simulated annealing (SA)\cite{753023}, have been proposed to iteratively search the optimal array. As pointed by Roux et al.\cite{7581110}, for large 2D sparse array design, SA performs much more efficient and robust than GA \cite{298712, hwang2006improving}. Initially, the cost function adopted by these works is a weighted combination of approximating desired continuous wave (CW) beam pattern (BP) (i.e., far-field narrowband BP) and  $l_1$ norm of element weighting coefficients. Later, optimizing multi-depth pressure field (e.g., 3 depths) strategy \cite{7549037,7581110} is introduced to better approximate and control the BP behavior of the sparse array in 3D space and the calculation of multi-depth pressure field can be accelerated \cite{6931817}. By simulating this realistic BP, new parameters, such as, excitation signal, element directivity and impulse response, even fabrication constraints can be included. 

Sparse arrays can also be divided into $on-grid$ and $out-of-grid$ ones. On-grid requires all the active elements should be placed on the intersections between equidistant rows and columns. In other words, elements in sparse arrays are selected from corresponding 2D dense arrays. In contrast, out-of-grid method poses no constraints on the element position distribution over the aperture excluding the overlapping. Theoretically, out-of-grid method allows better BP performance, (e.g., lower grating lobe levels). However, fabricating such kind of nongrid 2D sparse arrays requires more sophisticated and costly manufacture technique, e.g., capacitive micromachined ultrasound transducers (CMUTs) \cite{6217561, 8579867, 9815109}. 

Synthesis of sparse antenna arrays has always been a hot topic in antennas and wireless communication for decades \cite{WU2024100276}. Besides above mentioned stochastic optimization methods, sparse antenna arrays with complex excitation can also be designed through matrix pencil method \cite{4618700}, linear programming \cite{655623}, $l_1$ minimization\cite{5784319}, and reweighted $l_1$ minimization \cite{6193135, 6145602}. As pointed by Lebret et al. \cite{558465}, antenna array pattern synthesis problems with a limit on the number of active elements are usually nonconvex. Among these methods, linear programming, $l_1$ minimization, and weighted $l_1$ minimization belong to convex optimization, which reformulate the original nonconvex problems as convex problems through some assumptions or relaxions. 

Inspired by these pioneer works, we hereby formulate the 2D sparse array design problem with real-valued weights as a second-order cone programming (SOCP) and solve it iteratively using a re-weighted $l_1$ minimization technique \cite{6193135, 6145602}. Through this iterative optimization, a highly sparse 2D array can be designed with a BP strictly fulfilling certain constraints. This paper is organized as follows. Section II describes the SOCP formulation based on the assumption of a far-field narrowband BP. Section III presents a quasi-flattened side-lobe sparse (Q-Flats) array design method with SLL no higher than -21.26 dB and 252 activated elements and its simulation pipeline. Section IV compares BP characteristics, PSFs, and contrast across the Q-Flats and benchmark methods. Section V discusses results and makes concluding remarks. 

\section{Problem formulation}

\subsection{Far-Field Narrowband Beam Pattern (BP)}
Under the assumption of a far-field narrowband BP, each element is regarded as an omni-directional point source and transmits a monochromatic wave at the center frequency of an ultrasound array. Consider a planar array with elements uniformly spaced on an $N\times M$ dense grid (Fig. \ref{2D array.}), the array BP can be defined as \cite{655623, yang2021synthesis}
\begin{equation}
P(u,v) = \sum_{n=1}^{N}\sum_{m=1}^{M}w_{\substack{n,m}}e^{j\beta (n d_x u + m d_y v)}, \label{BP}
\end{equation}	
where $w_{\substack{n,m}}\in\mathbb{R}$ is the real-valued weight of each element for $n= 1,...,N$ and $m=1,...,M$; $\beta$ is the wavenumber; $d_x$ and $d_y$ denote the pitches of the elements in the $x$ (azimuth) and $y$ (elevation) directions, respectively; $u$ and $v$ are defined as $\sin(\theta)\cos(\varphi)-\sin(\theta_0)\cos(\varphi_0)$ and $\sin(\theta)\sin(\varphi)-\sin(\theta_0)\sin(\varphi_0)$, respectively; $(\theta_0, \varphi_0)$ represents the direction of beam steering (Fig. \ref{2D array.}). 

\subsection{Sparse Array Synthesis}  
The target 2D sparse array should have as few elements as possible while having a SLL that satisfies the prescribed BP constraints. The objective function for such a sparse array design can thus be formulated as
	\begin{subequations}\label{zeronorm}
	\begin{align}
	&\overset{}{\underset{\boldsymbol w}{\operatorname{min}}}\ \lVert \boldsymbol w \rVert_0 \label{target2a}\\
		\text {s.t. } &|P(u,v)|\leq D(u,v),\label{cond2c} \\
	     &P(u_0,v_0) =1, \label{cond2b}
	\end{align}
    \end{subequations}
where $\boldsymbol w  = [w_{\substack{1,1}},w_{\substack{1,2}},...,w_{\substack{n,m}}]^\top\in\mathbb{R}^{m\times n}_+$, $u_0 = u(\theta_0, \varphi_0)=0, v_0 = v(\theta_0, \varphi_0)=0$, and $D(u,v) $ is the upper limit or a mask on the BP and $u,v \in [-2,2]$.
 
Unfortunately, the optimization problem in Eq. \eqref{zeronorm} is non-convex and does not guarantee finding the global optimum. An alternative way to solve it is to relax it as the following convex optimization problem:
	\begin{subequations}\label{onenorm}
	\begin{align}
	&\overset{}{\underset{\boldsymbol w}{\operatorname{min}}}\ \lVert \boldsymbol w \rVert_1 \label{target3a}\\
		\text {s.t. } &|P(u,v)|\leq D(u,v),\label{cond3c} \\
         &P(u_0,v_0) =1. \label{cond3b}
	\end{align}
    \end{subequations}

The only difference between Eq.  \eqref{zeronorm} and Eq.  \eqref{onenorm} is the choice of the objective function. $l_1$ norm is well-known to produce sparse solutions in a wide range of applications, e.g., recovering sparse signals in compressed sensing \cite{1614066}. To further enhance the sparsity of the solution $\boldsymbol w$ given by problem \eqref{onenorm}, an iterative re-weighted $l_1$ minimization strategy \cite{lobo2007portfolio, candes2008enhancing} is exploited. It has been shown that such an strategy can approximate the original problem \eqref{zeronorm} more closely than only using unweighted $l_1$ norm, thus providing sparser solutions.

The weighted $l_1$ minimization problem to be solved at the $k_{th}$ iteration is written as follows
	\begin{subequations}\label{weightednorm}
	\begin{align}
		&\overset{}{\underset{\boldsymbol w_k}{\operatorname{min}}}\ \lVert {\boldsymbol c}_k^\top\boldsymbol w_k \rVert_1 \label{target4a}\\
		\text {s.t. } &|P(u,v)|\leq D(u,v),\label{cond4c} \\
		&P(u_0,v_0) =1, \label{cond4b}
	\end{align}
    \end{subequations} 
where $\boldsymbol w_k  = [w_{\substack{1,1}}^{k},w_{\substack{1,2}}^{k},...,w_{\substack{n,m}}^{k}]^\top\in\mathbb{R}^{m\times n}_+$, $\boldsymbol c_k  = [c_{\substack{1,1}}^{k},c_{\substack{1,2}}^{k},...,c_{\substack{n,m}}^{k}]^\top\in\mathbb{R}^{m\times n}_+$, and $c_{\substack{n,m}}^{k} = (|w_{\substack{n,m}}^{k-1}|+\epsilon)^{-1}$, $\epsilon>0$. Here the coefficients $\boldsymbol c_k$ used to weight the apodizations are positive and calculated on solution $\boldsymbol w_{k-1}$ from last step. Small weights $w_{\substack{n,m}}^{k-1}$ at previous step $k-1$ results in large weights $c_{\substack{n,m}}^{k}$ and are penalized at next step $k$, while large weights result in small weights that ensure their reconduction at the next iteration. The parameter $\epsilon$ is to stabilize the iterative process and ensures that a zero apodization $w_{\substack{n,m}}^{k-1}$ at last step allow $w_{\substack{n,m}}^{k} \neq 0$. As pointed by Candes et al. \cite{candes2008enhancing}, $\epsilon$ should be set slightly smaller than the non-zero magnitudes of the expected solution. Notably, at the first iteration ($k = 1$),  $\boldsymbol c_k  = \boldsymbol 1^\top$. In other words, this sequential optimization algorithm solves the problem \eqref{onenorm} in the beginning. The initialization and optimization process of this iterative re-weighted $l_1$ optimization algorithm are summarized in the Algorithm 1.

\begin{algorithm}[!t]
	\renewcommand{\algorithmicrequire}{\textbf{Input:}}
	\renewcommand{\algorithmicensure}{\textbf{Output:}}
	\newcommand{\INITIALIZATION}{\item[\textbf{1. Initialization:}]}
	\newcommand{\SEQUENTIAL}{\item[\textbf{2. Squential Re-weighted Optimization:}]}
	\caption{Re-weighted L1 SOCP}
	\label{alg1}
	\begin{algorithmic}
		\REQUIRE $M, N, \beta, d_x, d_y, \epsilon, and$ $D(u,v)$ 
		\ENSURE  $\boldsymbol w$
		\INITIALIZATION $\boldsymbol c_1 = \boldsymbol 1^\top$, $k=1$.
		\SEQUENTIAL
		\FOR {$k=1$ to L}
		\STATE \ding{192} Solve problem (4) to obtain $\boldsymbol w_k$;
		\STATE \ding{193} Normalize $\boldsymbol w_k$ with $\max{(\boldsymbol w_k)}$;
		\STATE \ding{194} Count the number of elements $N_{\substack{ele}}^{k}$ with weights $w_{\substack{n,m}}^{k}/\max{(\boldsymbol w_k)} \geq w_{thre}$;
		\STATE \ding{195} If $k$ $\geq 3 $ and $N_{\substack{ele}}^{k}$ == $N_{\substack{ele}}^{k-1}$ == $N_{\substack{ele}}^{k-2}$
		\STATE \hspace{1em}  $\boldsymbol {break}$;
		\STATE \hspace{1em}  End if
		\STATE \ding{196} $\boldsymbol c_{k+1} = (|\boldsymbol w_k|+\epsilon)^{-1}$. 
		\ENDFOR
		\RETURN  $\boldsymbol w$ $\leftarrow \boldsymbol w_k$ / $\max{(\boldsymbol w_k)}$
	\end{algorithmic}  
\end{algorithm}	

\subsection{Second-Order Cone Programming (SOCP)}
Although problem \eqref{weightednorm} is implicitly treated as convex optimization, its convexity still has not been verified until now. In this subsection, problem \eqref{weightednorm} is formulated as a SOCP which is explicitly a convex optimization. First, all the $w$ and $c$ are non-negative real numbers, and therefore, the objective function of problem \eqref{weightednorm} can be easily simplified as ${\boldsymbol c}_k^\top\boldsymbol w_k$, since $\lVert {\boldsymbol c}_k^\top\boldsymbol w_k \rVert_1 = \sum_{n=1}^{N}\sum_{m=1}^{M}|c_{\substack{n,m}}^{k}w_{\substack{n,m}}^{k}| =  \sum_{n=1}^{N}\sum_{m=1}^{M}c_{\substack{n,m}}^{k}w_{\substack{n,m}}^{k}$ = ${\boldsymbol c}_k^\top\boldsymbol w_k$. For the final constraints \eqref{cond4b}, $P(u_0,v_0) = P(0,0) = \sum_{n=1}^{N}\sum_{m=1}^{M}w_{\substack{n,m}} = 1$. For the first constraints \eqref{cond4c}, it can be reformulated as a second-order-cone (SOC) constraints as shown in \eqref{cond5c}. Based on these derivations, problem \eqref{weightednorm} can be expressed as
	\begin{subequations}\label{SOCP}
	\begin{align}
		&\overset{}{\underset{\boldsymbol w_k}{\operatorname{min}}}\ {\boldsymbol c}_k^\top\boldsymbol w_k  \label{target5a}\\
		\text {s.t. } &\lVert {\boldsymbol A(u,v)}\boldsymbol w_k \rVert_2\leq D(u,v),\label{cond5c} \\
		&\sum_{n=1}^{N}\sum_{m=1}^{M}w^k_{\substack{n,m}} =1, \label{cond5b}
	\end{align}
    \end{subequations}
where:
    \begin{equation}\label{dataseq}
    	\resizebox{1\hsize}{!}{$
    	\begin{aligned}    		
    		\boldsymbol A(u,v) & =\left[\begin{array}{cccc}
    		\cos(\beta (x_{\substack{1,1}} u + y_{\substack{1,1}} v)), & \sin(\beta (x_{\substack{1,1}} u + y_{\substack{1,1}} v)) \\
    		\cos(\beta (x_{\substack{1,2}} u + y_{\substack{1,2}} v)), & \sin(\beta (x_{\substack{1,2}} u + y_{\substack{1,2}} v)) \\
    			... & ... \\
    			... & ... \\
    	    \cos(\beta (x_{\substack{m,n}} u + y_{\substack{m,n}} v)), & \sin(\beta (x_{\substack{m,n}} u + y_{\substack{m,n}} v)) \\
    		\end{array}\right] ^\top, \\
    		\boldsymbol w_k  &=\left[\begin{array}{llll}
    		w_{\substack{1,1}}^{k},w_{\substack{1,2}}^{k},...,w_{\substack{n,m}}^{k}
    		\end{array}\right] ^\top\in\mathbb{R}^{m\times n}_+, \nonumber\\
    		\ u,v  &=\left[\begin{array}{cccc}
    			-2,2
    	    \end{array}\right].		
    	\end{aligned}$}       
    \end{equation}
Now problem \eqref{SOCP} is a standard SOCP \cite{boyd2004convex}. The objective function \eqref{target5a} is a linear function. The constraints \eqref{cond5c} and \eqref{cond5b} are SOC, and linear equality constraints, respectively. To finally solve the problem \eqref{SOCP}, the constraint \eqref{cond5c}, which is parameterized by $u,v$, are approximated by discretizing $u,v$ evenly at a certain density. Low density cannot completely control the BP. High density can increase the computational accuracy, but the number of SOC constraints increases quadratically for designing 2D arrays, therefore computational burden dramatically.

\begin{figure}[!t]
	\centering
	\includegraphics[width=\columnwidth]{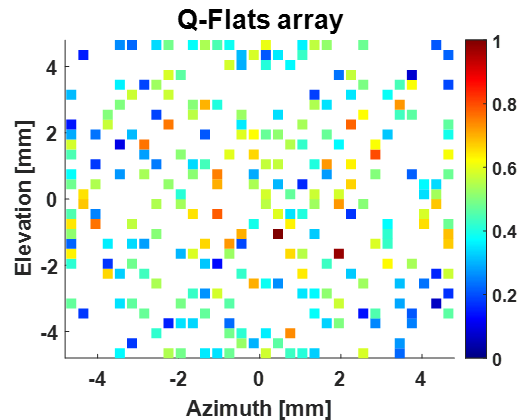}
	\caption{The layout of the optimized quasi-flatten side-lobe (Q-flats) sparse array (252 activated elements). Each square denotes a single activated elements in corresponding dense array and the color denotes the normalized apodization.}
	\label{array_arrangement}
\end{figure}

\section{Methods}
\subsection{2D Sparse Arrays Design}
\label{sec:Methods}

The dense matrix array M3dV is used as reference array in this paper, which has 1024 elements in a $32\times 32$ grid without any blanks. The main parameters of M3dV is shown in Table I. Based on the grided layout of the M3dV matrix array, a 2D sparse array with maximum SLL lower than a certain value is designed. Once the main-beam width is determined, the maximum SLL is suppressed as low as possible until the number of active elements in the finally obtained array exceeds 256.

\begin{figure}[!t]
	\centering
	\includegraphics[width=\columnwidth]{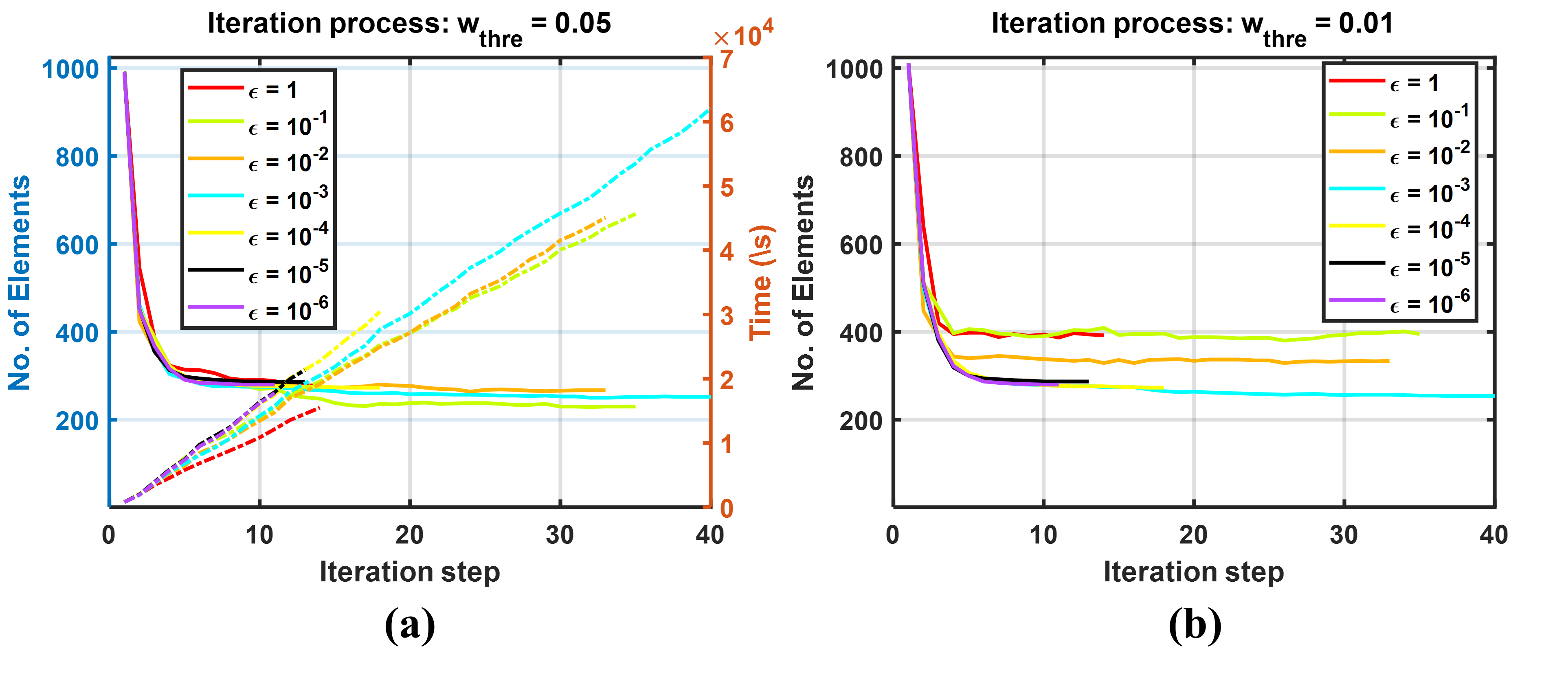}
	\caption{The number of activated array elements and accumulated computational time versus iteration steps with different $\epsilon$ values and $w_{thre}$ (a) 0.05, and (b) 0.01.}
	\label{iteration_process}
\end{figure}

Through trials and errors, a 2D sparse array was designed through the optimization problem \eqref{SOCP} by using Algorithm 1. The values of $M$, and $N$ are both 32. The mask $D(u,v)$ of BP is designed as 
\begin{equation}\label{SLL_21dot2}
          D(u,v) = 0.0865, \quad  0.055 ^{2} \leq u^2 + v^2 \leq [1+\sin(41^{\circ})]^2. 	
\end{equation}	
Here $20*\lg(0.0865)$ equals about -21.26 dB. The 0.055 indicates the radius of the main-lobe region is no more than 0.055. When discretizing problem \eqref{SOCP}, the value of $\Delta u$ and $\Delta v$ are both 0.005. The parameters used to calculate BP is from Table I. The $\epsilon$ and $w_{thre}$ is set as 0.001 and 0.05, respectively. The problem \eqref{SOCP} is solved by the Yalmip toolbox \cite{lofberg2004yalmip} with Mosek \cite{mosek2015mosek} solvers in MATLAB (Version 2019b, Mathworks Inc. MA, USA). Finally, a 2D sparse array with maximum SLL of -21.26 dB and 252 activated elements was obtained after 42 iterations. The positions and apodizations of all the activated elements are shown in Fig. \ref{array_arrangement}. The number of active elements in each iteration is shown in Fig. \ref{iteration_process}. The dense array-M3dV and our designed sparse array are hereinafter referred as Dense and Q-Flats, respectively. 

Besides the designed Q-Flats, a Fermat spiral array, and a density-tapered spiral array according to a $50\%$-Tukey window \cite{8832242} were also implemented on the grided layout of the Dense as comparison. They are hereinafter referred as Spiral and Spiral-Taper, respectively. The elements belonging to Spiral were selected among those of the Dense, by activating the on-grid elements whose positions are closet to the ideal positions of an ungrided, 9.6-mm-wide with 256 seeds spiral array. The element arrangements of the Spiral-Taper were also on the layout of the Dense and in the same way as designing the Spiral.     

\begin{table}
	\centering
	\caption{Parameters of matrix array M3dV}
	\label{tab1}
	\begin{tabular}{cc}
		\toprule
		$\textbf{Parameters}$ & $\textbf{Values}$ \\ \hline
		Center frequency [MHz] & 3.0 \\ 
		Number of elements  & $1024 (32\times32)$ \\ 
		Pitch [mm] & $0.3$\\
		Elevation Aperture [mm] & $9.6 \times 9.6$ \\
		\toprule 
	\end{tabular}
	\vspace{-5mm}
\end{table}

\subsection{Field II Simulation Setup}
The imaging performance of the four arrays (Dense, Sparse, Spiral, and Spiral-Taper) were simulated using Field II \cite{139123, jensen1997field} in MATLAB (2019 b). Since all four arrays were realized on the same matrix array layout-M3dV array, a $32\times 32$ 2D matrix array transmitting a Hanning modulated 1-cycle sinusoidal pulse with a central frequency of 3 MHz and 60$\%$ bandwidth were simulated in this work. The transducer parameters were the same as those used in 2D Sparse array design and detailed in Table \ref{tab1}. The same steered diverging wave (DW) imaging sequence \cite{roux2018experimental,bernal2020high} was applied to the four arrays. 49 volumes were acquired by using $7\times 7$ virtual sources distributed over a spherical cap at about 8.05 mm away from the center of the array. All the elements were used to transmit and receive, resulting an opening angle of $60^{\circ}\times 60^{\circ}$. The distribution of steering angles were from - $30^{\circ}$ to $30^{\circ}$ at a $10^{\circ}$ step in both azimuth and elevation directions. Pyramidal volumes were beamformed with delay and sum (DAS) algorithm for each DW transmission and subsequently coherently compounded to form a final high-quality volumetric image with $60^{\circ}\times 60^{\circ}$ opening angles both in the azimuth and elevation dimensions. The speed of sound in all simulations was set to 1540 m/s. The sampling frequency was 100 MHz.

To evaluate the point spread functions (PSFs) of the four different 3D imaging system, two numeric phantoms were developed. For the first phantom, four scatters were placed on the axial axis at 20, 50, 80, and 110 mm depth, respectively, to assess the on-axis PSFs at different depths. For the second phantom, four scatters were placed on the off-axial axis with an azimuth angle of $13^{\circ}$ and another four scatters with an elevation angle of $13^{\circ}$. 

To assess the contrast performance, a tissue mimicking phantom was generated by positioning point scatters randomly in a cube of 50 mm $\times$ 50 mm $\times$ 20 mm (azimuth $\times$ elevation $\times$ axial) centered at a depth of 50 mm. The density of scatters was $16/\text{mm}^{3}$ to produce fully developed speckle. Five 10-mm-diameter spheroidal cystic regions were placed inside the cubic phantom and centered at (0, 0, 50), (-15, 0, 50), (15, 0, 50), (0, -15, 50), and (0, 15, 50) mm, respectively. The scattering amplitude of scatters inside these five cystic reigons were set as 0 while Gaussian distributed amplitudes were assigned to the remaining scattering inside the phantom. 

\subsection{Performance Metrics}
Spatial resolutions (e.g., azimuthal, and elevational resolution) were evaluated by measuring the full-width at half-maximum (FWHM) from the $C$ plane cross each of scatters. Moreover, to predict the contrast performance of the four arrays, the mean-sidelobe-level (MSLL) was calculated by averaging the gray level between the -80 dB and -6 dB isoline in the $C$ plane.

For cyst regions, the contrast ratio (CR) and generalized contrast-to-noise ratio (gCNR) were calculated to assess the lesion detectability of the four arrays, as follows:

\begin{equation}
	\boldsymbol {CR} = 20 \ast log_{10}(\frac{\mu_{cyst}}{\mu_{back}}). \label{CR}
\end{equation}	

\begin{equation}
	\boldsymbol {gCNR} = 1-\int_{-\infty}^{+\infty}{\underset{\boldsymbol x}{\operatorname{min}}}\ ({p_{back}(x), p_{cyst}(x)})dx. \label{gCNR}
\end{equation}	
Where $\mu_{cyst}$, is the mean amplitude of the region of interest (ROI) inside the cyst, $\mu_{back}$ is the mean amplitude of the ROI in background tissues. The $p_{back}(x)$, and $p_{cyst}(x)$ are the probability density function of values taken by the cyst ROI pixels and background ROI pixels, respectively.

\section{Results}

\begin{figure*}[!t]
	\centering
	\includegraphics[width=\textwidth]{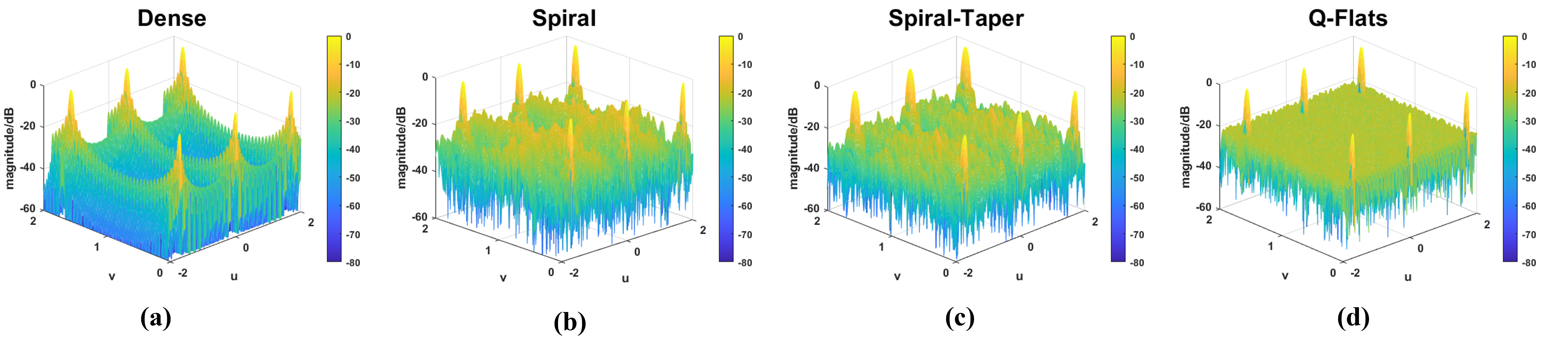}
	\caption{Normalized magnitude in dB of 2D beam pattern (BP) of the (a) Dense, (b) Spiral, (c)Spiral-Taper, (d) Q-Flats arrays, respectively.}
	\label{2D BP}
\end{figure*}

\begin{figure*}[!t]
	\centering
	\includegraphics[width=\textwidth]{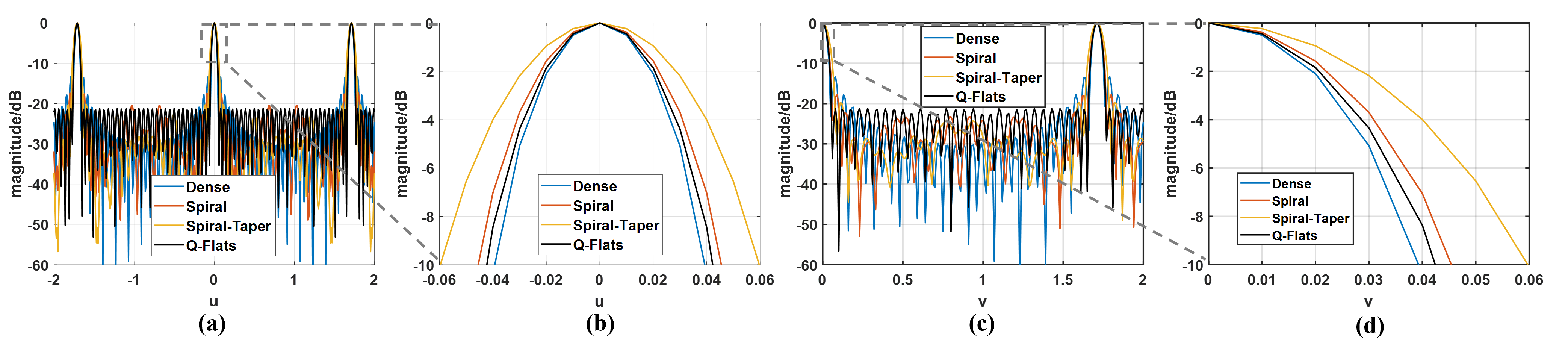}
	\caption{Comparison of normalized magnitude in dB of 2D BP of the Dense, Spiral, Spiral-Taper(50$\%$-Tukey), and Q-Flats arrays at (a) v = 0 plane (c) u = 0 plane. (b) and (d) are zoomed-in main-lobe regions within -10 dB indicated by gray-dashed line in (a) and (c), respectively. Here, $u$=$\sin(\theta)\cos(\varphi)-\sin(\theta_0)\cos(\varphi_0)$, $v$ =  $\sin(\theta)\sin(\varphi)-\sin(\theta_0)\sin(\varphi_0)$. $(\theta_0, \varphi_0)$ represents the beam steering direction (Fig. \ref{2D array.})}
	\label{v=0}
\end{figure*}

\begin{figure*}[ht]
	\centering
	\includegraphics[width=0.7\textwidth]{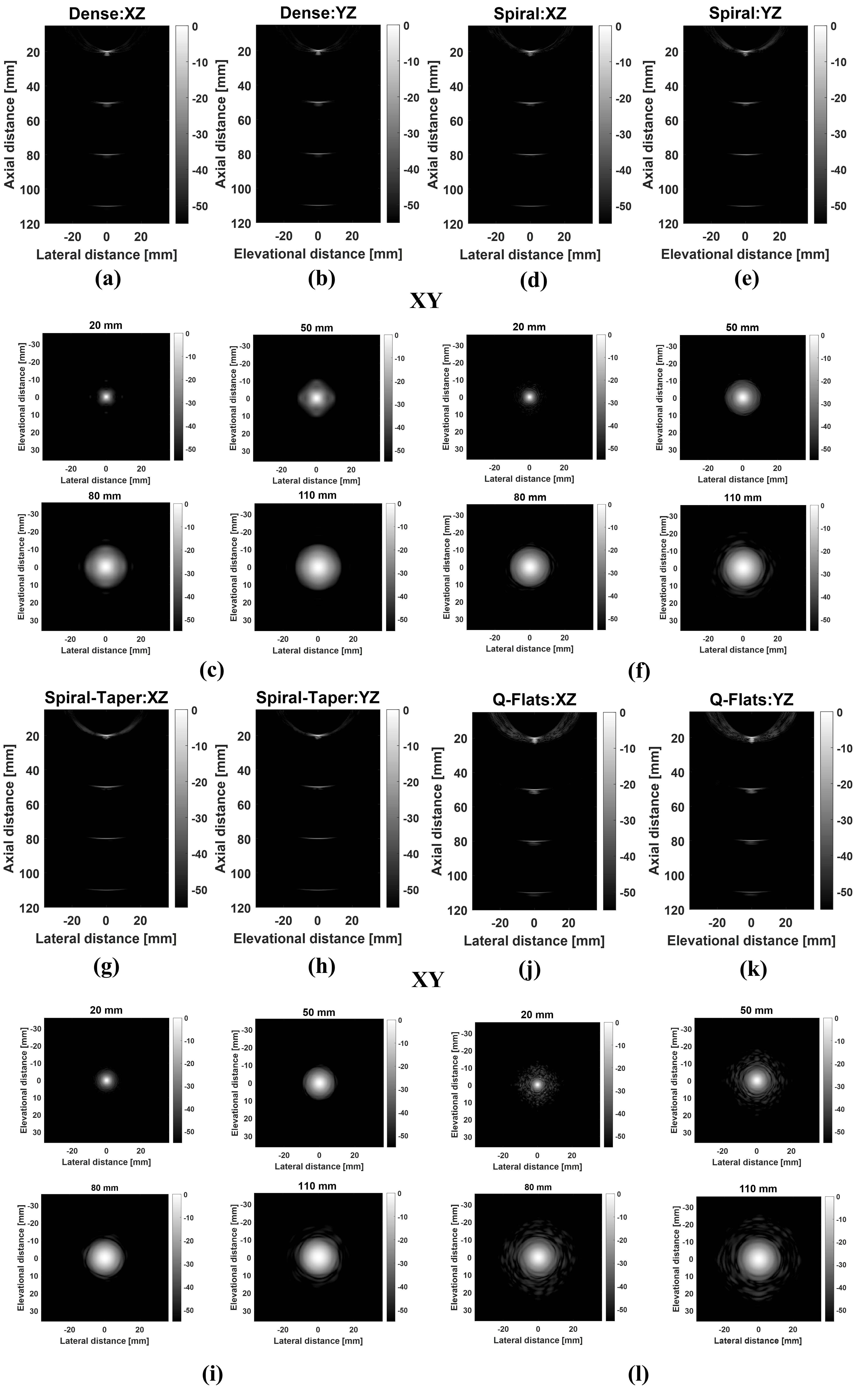}
	\caption{Simulated on-axis PSFs using various arrays including Dense (a)-(c), Spiral (d)-(f), Spiral-Taper (g)-(i), and Q-Flats (j-L) . XZ: azimuth-axial plane. YZ: elevation-axial plane. XY: $C$ planes at four depths of the point scatters. }
	\label{point_bmode}
\end{figure*}

\subsection{The effect of parameters $\epsilon$ and $w_{thre}$}
Fig. \ref{iteration_process} shows the number of activated elements and the total computational time as a function of iteration step with different values of $\epsilon$ and $w_{thre}$. When $\epsilon$ equals to 0.001, the sparsest array was achieved with a number of activated elements of 252, a total iteration count of 40, and a total computational time of about 17 hours. In general, about 60$\%$ elements were reduced within the first three iterations and the total computation time scaled linearly with the total number of iterations. The time cost of one iteration was no more than about 28 minutes on average despite a small variation with respect to the $\epsilon$. 

As is shown in Fig. \ref{iteration_process} (a), when $\epsilon$ was equal to 0.1, the iteration processes achieved a sparse array with less activated elements of 230 and less iteration step of 35. However, as indicated by Fig. \ref{iteration_process} (b), when $\epsilon$ was larger than 0.001, hundreds of element weights fell between 0.01 and 0.05 after normalization, which caused a significant deviation from desired BP. When $\epsilon$ was smaller than 0.001, only several elements were between 0.01 and 0.05. Finally, it seemed that excessive values of $\epsilon$ tended to result in less sparse solutions than overly-small ones. To sum up, the most suitable value for $\epsilon$ in our case was 0.001, which achieved a truly sparsest array.

\subsection{Theoretical Analysis of Far-Field Beam Pattern (BP)}

Figs. \ref{2D BP} (a)-(d) show the normalized magnitude 2D far-field BP of the four arrays-Dense, Spiral, Spiral-Taper, and Q-Flats, respectively. Only half of the amplitude of BP ($-2 \leq u \leq 2, 0\leq v \leq 2$) was plotted due to symmetry properties. In general, all the four BPs had one main lobe at $ u = 0,v = 0$ and five grating lobes with the same amplitude at $ u = 1.72, v = 0; u = -1.72, v = 0; u = 1.72, v = 1.72; u = -1.72, v = 1.72; u = 0, v = 1.72$, respectively. The BP of the Dense had more regions with magnitude of about -60 dB than those of Spiral, Spiral-Taper, and Q-Flats since the Dense had much more activated elements (1024). The BP of the Q-Flats had quasi-flatten SLLs no more than -21.26 dB, which strictly satisfied the predetermined constraints on side-lobe regions. 

Figs. \ref{v=0} (a) and (c) show the BP profiles of the four arrays in $v =0$ plane and $u=0$ plane, respectively. Figs. \ref{v=0} (b) and (d) show zoomed-in main-lobe profiles of the four arrays as outlined by the gray dashed box in (a) and (c). The Dense had the narrowest main-lobe width and the Spiral-Taper the widest main-lobe width both in $v =0$ plane and $u=0$ plane among the four arrays. Since the Spiral-Taper had the smallest aperture size. Besides, the Q-Flats had a little narrower main-lobe width compared with that of the Spiral both in $v =0$ plane and $u=0$ plane. 

For the SLLs, as is shown in Figs. \ref{v=0} (a) and (c), the Dense and Spiral-Taper had relatively low SLLs. However, the Dense had the highest first side lobe of about -13 dB close to the main lobe. The Q-Flats had the highest SLLs visually, but it has the lowest first side lobe of about -21.26 dB among the four arrays.

\subsection{ Simulated Point Phantom}
Fig. \ref{point_bmode} shows the simulated PSFs of on-axis ($0^{\circ}$ for both elevation and azimuth directions) scatters obtained by the four arrays-Dense, Spiral, Spiral-Taper, and Q-Flats at different depths of 20, 50, 80, and 110 mm, respectively. For each array, all the azimuth-axial plane (XZ), elevation-axial plane (YZ), and four $C$ planes (XY) of scatters at different depths were presented. Table \ref{tab2} summarizes the FWHM and MSLL values of these on-axis PSFs. Visually, resolution of the four arrays decreased significantly as the depth of point scatters increased. In near-field region (20 mm in depth), except for the Dense, the remaining three arrays showed obvious side lobes. The Dense and Q-Flats presented a little stronger axial lobes compared with the Spiral and Spiral-Taper, which might be caused by edge waves from edge elements \cite{7103534}. In the $C$ planes, compared with other PSFs, the Q-Flats showed more obvious sporadic side lobes in edge regions, however, fortunately, the levels of these side lobes were almost lower than -40 dB. 
   
Quantitatively, as shown in Table \ref{tab2}, the Dense performed the best and the Spiral-Taper worst in terms of resolution among the four arrays regardless of depth. Since the Spiral-Taper had the smallest effective aperture size. The Q-Flats showed a little better resolution ($\geq 3\%$) than the Spiral especially in deep regions, which was consistent with the BP results in Fig. \ref{v=0}. Finally, the four arrays performed the same in both the XZ and YZ planes. For the MSLL, similar to resolution, the Dense, and the Spiral-Taper had the lowest and highest MSLL, respectively. The Spiral had a little lower MSLL than the Q-Flats ($\leq 0.97 dB$). It should be noticed that at a depth of 110 mm, the MSLL of Q-Flats was lower than that of the Spiral about 0.45 dB exceptionally.    

\begin{figure}[!t]
	\centering
	\includegraphics[width=\columnwidth]{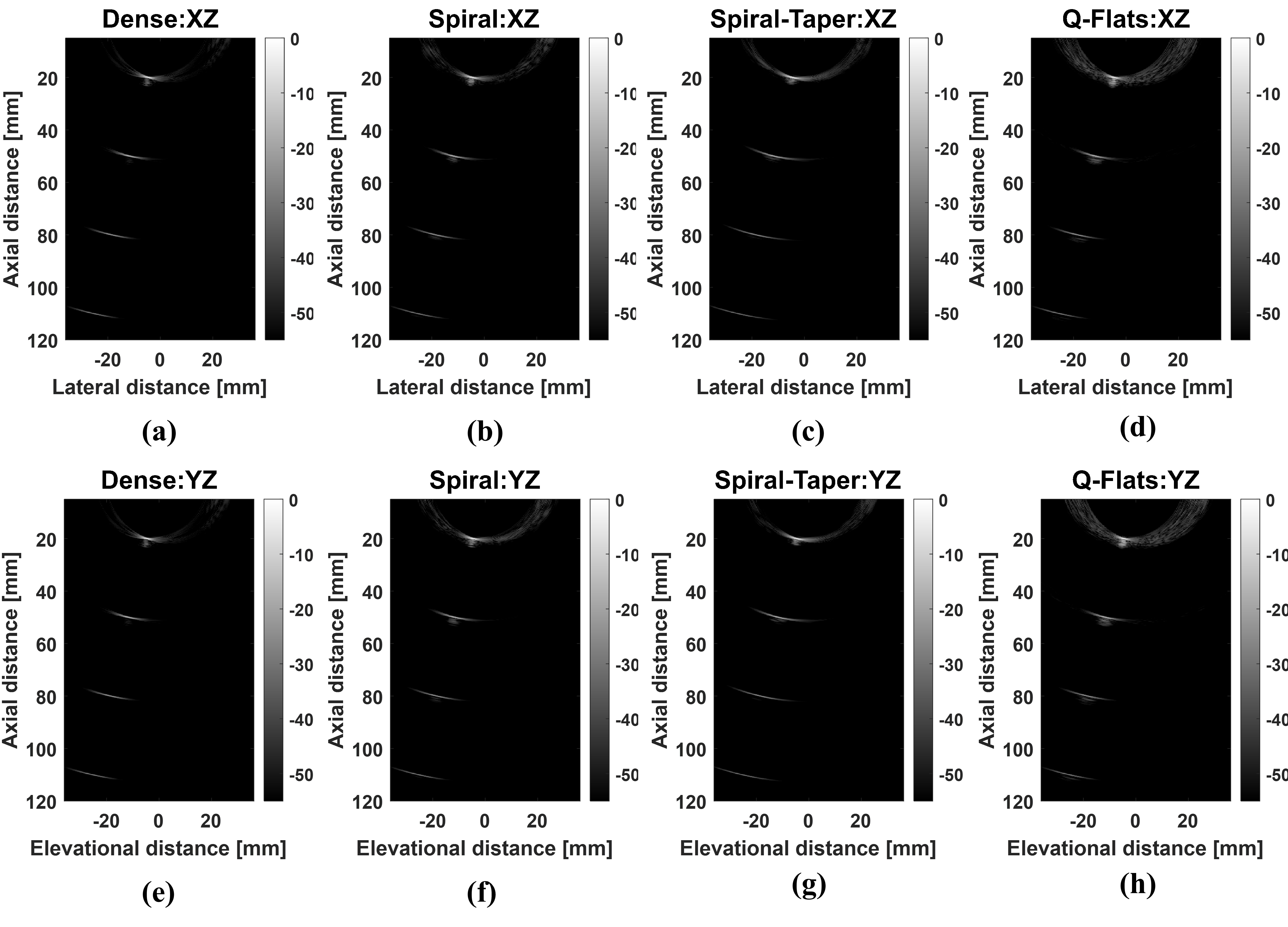}
	\caption{Simulated off-axis ($13^{\circ}$ for elevation or azimuth direction) PSFs using various arrays including Dense (a) and (b), Spiral (c) and (d), Spiral-Taper (e) and (f), and Q-Flats (g) and (h). XZ: azimuth-axial plane. YZ: elevation-axial plane.}
	\label{steered_point_bmode}
\end{figure}

Fig. \ref{steered_point_bmode} shows the simulated PSFs of off-axis ($13^{\circ}$ for elevation or azimuth direction) scatters obtained by the Dense, Spiral, Spiral-Taper, and Q-Flats at different depths of 20, 50, 80, and 110 mm, respectively. In general, all the arrays exhibited comparable steering performance with no grating lobes visually. Compared with the on-axis results in Fig. \ref{point_bmode}, the Dense showed no axial lobes anymore at depth of 50 mm, but the Spiral presented more obvious axial lobe at the same depth.

\subsection{ Simulated Cyst Phantom}
\begin{figure*}[!t]
	\centering
	\includegraphics[width=1.0\textwidth]{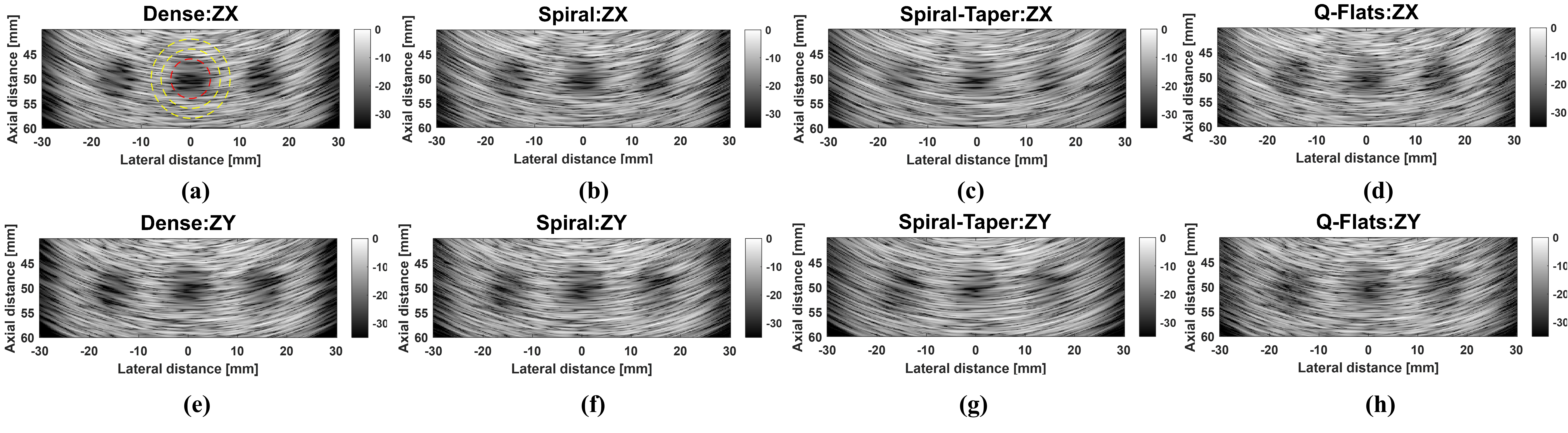}
	\caption{Simulated B-mode images of the cyst phantom using various arrays including Dense (a) and (e), Spiral (b) and (f), Spiral-Taper (c) and (g), and Q-Flats (d) and (h). The top row for XZ slices, and the bottom row for YZ slices. XZ: azimuth-axial plane. YZ: elevation-axial plane.}
	\label{cyst_bmode}
\end{figure*}

Fig. \ref{cyst_bmode} presents the simulated B-mode images of the cyst phantom by using the Dense, Spiral, Spiral-Taper, and Q-Flats. Visually, the Dense showed the best performance in terms of contrast and cysts boundary delineation. And the Spiral-Taper performed the worst, especially for the cysts located on off-axis positions. The Spiral performed slightly better than the Q-Flats mainly in contrast performance. 

Table \ref{tab3} lists the calculated CR and gCNR values. As shown in Fig. \ref{cyst_bmode} (a), the area inside the red dashed line with a radius of 4 mm was taken as cyst region, while the region inside the two concentric yellow dashed lines was selected as the backgroud region. Quantitatively, the Dense, and the Spiral-Taper had the highest and lowest CR and gCNR values on average, respectively. The CR value of the Spiral was slightly higher than that of the Q-Flats averagely (0.3 dB). The Q-Flats had the same gCNR value on average with the Spiral. All these quantitative results were consistent with the visual performance.

\begin{table*}[!t]
	\centering
	\caption{The Full Width at Half Maximum (FWHM [mm]) and Mean Side-Lobe Level (MSLL [{d}B]) for The Four Different Arrays.}
	\label{tab2}
	\begin{adjustbox}{width=\textwidth}
		\begin{tabular}{|c|c|c|c|c|c|c|c|c|c|c|c|c|}
			\hline
			\multirow{3}{*}{\textbf{Array}} & \multicolumn{8}{c|}{\textbf{FWHM [mm]}} &\multicolumn{4}{c|}{\multirow{2}{*}{\textbf{MSLL [dB]}}}\\ \cline{2-9}
			
			\multirow{3}{*}{} & \multicolumn{4}{c|}{\textbf{XZ}} &\multicolumn{4}{c|}{\textbf{YZ}} & \multicolumn{4}{c|}{\multirow{2}{*}{}} \\ \cline{2-13}
			
			\multirow{3}{*}{} & \textbf{20 [mm]} & \textbf{50 [mm]} & \textbf{80 [mm]} & \textbf{110 [mm]} & \textbf{20 [mm]} & \textbf{50 [mm]} & \textbf{80 [mm]} & \textbf{110 [mm]} & \textbf{20 [mm]} & \textbf{50 [mm]} & \textbf{80 [mm]} & \textbf{110 [mm]} \\ \hline
			
			\textbf{Dense} & 1.38 (+0.0\%)& 3.00 (+0.0\%) & 4.74 (+0.0\%)& 6.06 (+0.0\%) & 1.38 (+0.0\%) & 3.00 (+0.0\%)& 4.74 (+0.0\%)& 6.06 (+0.0\%)& -45.22 & -38.69 & -36.83 & -36.74 \\ \hline
			
			\textbf{Spiral} & 1.50 (+8.7\%) & 3.30 (+10.0\%)& 5.19 (+9.5\%)& 6.42 (+5.9\%)& 1.50 (+8.7\%)& 3.30 (+10.0\%)& 5.19 (+9.5\%)& 6.42 (+5.9\%)& -48.46 & -39.29 & -36.65 & -36.04 \\ \hline
			
			\textbf{Spiral-Taper} &  1.71 (+23.9\%) & 3.99 (+33.0\%) & 6.24 (+31.7\%) & 7.26 (+19.8\%)& 1.71 (+23.9\%)& 3.99 (+33.0\%)& 6.24 (+31.7\%)& 7.26 (+19.8\%)& -44.60 & -36.34 & -34.82 & -35.39 \\ \hline
			
			\textbf{Q-Flats} & 1.50 (+8.7\%) & 3.21 (+7.0\%)& 4.98 (+5.1\%) & 6.24 (+3.0\%)& 1.50 (+8.7\%)& 3.21 (+7.0\%)& 4.98 (+5.1\%)& 6.24 (+3.0\%)& -47.75 & -38.32 & -35.80 & -36.49  \\ \hline
			
		\end{tabular}
	\end{adjustbox}
\end{table*}

\section{Conclusion and Discussion}
\label{sec:guidelines}
\subsection{Brief summaries}
In this study, we introduced an alternative way to synthesis 2D sparse array through sequential convex optimization. Specifically, the original non-convex optimization problem for array synthesis was relaxed and  formulated as a iteratively-solved SOCP through re-weighting technique. To the best of our knowledge, we were the first one who conducted this method on 3D medical ultrasound imaging. Based on this method, a 252-elements sparse array with SLL no more than -21.26 dB - Q-Flats was designed based on the layout of a $32\times 32$ dense array. The imaging performance of the Q-Flats array was compared with a dense array (Dense), a 256-elements uniform spiral array (Spiral), and a 50 $\%$-Tukey window-tapered 256-elements spiral array (Spiral-Taper). Simulated PSFs (Fig. \ref{point_bmode} and \ref{steered_point_bmode}) and cyst phantom results (Fig. \ref{cyst_bmode}) showed that as expected, the Dense performed the best in terms of resolution and contrast. The Q-Flats showed slightly better resolution and worse contrast than Spiral, which seemed to provide a good compromise between resolution and contrast. The Spiral-Taper performed the worst in both the resolution and contrast. Theoretically, the Spiral-Taper tended to show better contrast than the Spiral at the sacrifice of resolution. However, based on the simulation results both visually and quantitatively, the contrast performance of Spiral-Taper was significantly worse than the Spiral. The possible explanations were as follows: first, since the Spiral-Taper was realized by discretizing the ideal tapered spiral array on a dense array as described in Section \ref{sec:Methods}, the deviation of element positions caused by the discretization to its ideal position might influence the performance of the Spiral-Taper; Second, the excessively low resolution of the Spiral-Taper severely influenced the detection and delineation of cysts, which hinted that for small size matrix array imaging, main-lobe width might play a more important role than SLL. Besides, the results of the Spiral-Taper also indicated that randomly selected sparse array might deteriorate the imaging resolution and performance simultaneously, which underlined the significance of designing sparse array by appropriate methods (e.g., optimization-based methods). Finally, all the simulated results showed no grating lobes because the maximum steering angle for diverging wave transmission in this work is $30^{\circ}$, and the theoretically-allowed maximum steering angle was $41^{\circ}$ under the far-field narrow-band BP assumption.  

For the re-weighted L1 techniques, as shown in Fig. \ref{iteration_process}, the hyper-parameter $\epsilon$ had a significant influence on the iteration process. An appropriate $\epsilon$ could achieve a sparser result within a moderate total iteration steps. Besides, the variation of computation time for each step and $\epsilon$ was relatively small. The computation time for each step mainly depended on the scale of problems-the number of optimization variables and SOC constraints. Finally, due to the non-stochastic nature of the optimization process, compared with the stochastic optimization methods such as GA and SA, the final solution of the re-weighted L1 method was significantly more stable and reliable. 

\subsection{Feasible and Infeasible Problems}
In this work, a sequential convex optimization problem was solved iteratively (e.g., \eqref{weightednorm}). During the iterative process, the constraints \eqref{cond4c} and \eqref{cond4b} needed to be satisfied remaining unchanged and only the objective function was updated based on the previous step. Therefore, the feasibility of this series of convex problems was the same. At the first iteration (problem \eqref{onenorm}), the value of the objective function was constantly equal to 1, which was determined by equality constraint function \eqref{cond3b}. Thus, the feasibility of this sequential convex optimization problem could be verified at the very beginning. The feasibility illustrated that the predetermined BP constraints could be reached with at least one feasible point. 

Infeasibility indicated that the predetermined BP constraints could not be reached with any apodizations. In practical design, the infeasibility was usually caused by three type of reasons: excessively narrow main lobe, or excessively low SLL, or both. Excessively narrow main lobe meant that main-lobe width narrower than certain values could not be reached. For instance, if the radius of main-lobe region was smaller than 0.03 with SLLs lower than -13 dB, the problem was infeasible. Excessively low SLL meant that under certain appropriate main-lobe width, SLL lower than certain levels could not be reached. For instance, in our case, the radius of main-lobe region is set as 0.055 in \eqref{SLL_21dot2}, if the SLL is set lower than -30 dB, the problem becomes infeasible. Thus, one of the possible advantage of formulating the sparse array synthesis problem as a SOCP was that the achievement of a desired BP could be verified compared with those adopting stochastic optimization methods.  

\begin{table*}[!t]
	\centering
	\caption{The Contrast Ratio (CR [\lowercase{D}B]) and Generalized Contrast-to-Noise Ratio (\lowercase{G}CNR) for The Four Different Arrays.}
	\label{tab3}
	\begin{adjustbox}{width=\textwidth}
		\begin{tabular}{|c|c|c|c|c|c|c|c|c|c|c|c|c|c|c|}
			\hline
			\multirow{3}{*}{\textbf{Array}} & \multicolumn{6}{c|}{\textbf{CR [dB]}} &\multicolumn{1}{c|}{\multirow{3}{*}{\textbf{Avg. of CR}}} & \multicolumn{6}{c|}{\textbf{gCNR}} &\multicolumn{1}{c|}{\multirow{3}{*}{\textbf{Avg. of gCNR}}}\\ \cline{2-7} \cline{9-14}
			
			\multirow{3}{*}{} & \multicolumn{3}{c|}{\textbf{XZ}} &\multicolumn{3}{c|}{\textbf{YZ}} &\multirow{3}{*}{} & \multicolumn{3}{c|}{\textbf{XZ}} &\multicolumn{3}{c|}{\textbf{YZ}} &\multicolumn{1}{c|}{\multirow{3}{*}{}} \\ \cline{2-7} \cline{9-14}
			
			\multirow{3}{*}{} & \textbf{Left} & \textbf{Middle} & \textbf{Right} & \textbf{Left} & \textbf{Middle} & \textbf{Right} &\multirow{3}{*}{} & \textbf{Left} & \textbf{Middle} & \textbf{Right} & \textbf{Left} & \textbf{Middle} & \textbf{Right} &\multirow{3}{*}{} \\ \hline
			
			\textbf{Dense} & -8.83 & -9.08 & -8.78 & -9.07 & -9.40 & -9.25 & -9.07 & 0.64 & 0.63 & 0.57 & 0.63 & 0.64 & 0.65 & 0.63\\ \hline
			
			\textbf{Spiral} & -7.77 & -8.09 & -6.98 & -8.17 & -8.64 & -8.18 & -7.97 & 0.57 & 0.58 & 0.50 & 0.59 & 0.61 & 0.59 & 0.57\\ \hline
			
			\textbf{Spiral-Taper} & -5.85 & -6.04 & -4.61 & -6.96 & -6.96 & -6.41 & -6.14 & 0.47 & 0.48 & 0.39 & 0.51 & 0.51 & 0.48 &0.47 \\ \hline
			
			\textbf{Q-Flats} & -7.45 & -7.36 & -7.46 & -7.60 & -7.91 & -8.25 & -7.67 & 0.55 & 0.56 & 0.52 & 0.56 & 0.59 & 0.62 & 0.57 \\ \hline
			
		\end{tabular}
	\end{adjustbox}
\end{table*}

\subsection{With and Without Reweighted L1 Algorithm}
The original non-convex problem \eqref{zeronorm} was relaxed to convex problem \eqref{onenorm} by replacing the $l_0$ norm with $l_1$ norm in the objective function. Without using the re-weighted L1 algorithm (only one iteration), as shown in Fig. \ref{iteration_process}, regardless of the parameter $\epsilon$, the number of elements always equalled to 992. This showed that simply using $l_1$ norm to approximate $l_0$ norm was far from enough in reducing the extra elements as much as possible. Through adopting the re-weighted technique, the number of activated elements was quickly reduced by about 60$\%$ at the first four iterations. Then, the number of activated elements decreased gradually until meeting the stopping criterion in Algorithm 1. In summary, adopting the re-weighted technique could significantly and effectively thin a dense array.

The threshold used to determine which elements should be activated was no less than five percent of the maximum weight in Algorithm 1. A high threshold (e.g., 0.1) can reduce more elements, but it can possibly filter out elements with relatively high weights leading to BP deviation to the desired one. A low threshold (e.g., 0.01) rejected less elements, but it might include extra elements with very low weights.

\subsection{On-grid and Out-of-grid Array Layout}
\label{formats}

The re-weighted L1 SOCP method used in this work was appropriate for designing on-grid sparse arrays. As mentioned in Sec \ref{sec:Methods}, sparse array design in this work was based on the $32\times 32$ grided layout of the M3dV matrix array with element pitch equal to 0.3 mm. Another on-grid design method was the Costas array method, which was not compared in this work. If a 256-element sparse array was designed by the Costas array method, the aperture of M3dV array should be first evenly divided into a $256\times 256$ grid with an element pitch of 0.3/8 = 0.0375 mm and then one layout with 256 elements from 65534 different Costas arrays based on some criteria (e.g., peak side-to-main lobe ratio) was selected. Since one of the aim of this study was to quickly implement the designed sparse array on a mass-produced 2D dense array, it seemed that the re-weighted L1 SOCP method was more suitable than the Costas array method in this case. Besides, the re-weighted L1 SOCP method could also be easily extended to a $256\times 256$ grid case theoretically. 

Spiral arrays belonged to out-of-grid array layouts and were realized on the M3dV dense array in this work. It should be pointed that out-of-grid array architectures could be seen as selecting a certain number of elements (e.g., 256) from a grid with infinitely small grid size. Practically, in Matlab, numbers were recorded by using double-precision floating-point data type, which was finite precision. Therefore, spiral arrays could also be treated as on-grid ones in some sense.

\subsection{Side lobes and Grating lobes}
Narrow-band far-field BP was adopted to predict each kind of array performance on the magnitude and positions of the side lobes and grating lobes in this work. As shown in Fig. \ref{2D BP}, although the BPs of the four arrays showed different characteristics, they all had five grating lobes at the same positions as exact replica of their respective main lobes. Because the positions and amplitudes of grating lobes could not be controlled. Only the grating lobe width and SLLs could be adjusted. This could be explained as follows. For simplicity, consider the case where 
$v = 0$, \eqref{BP} could be written as

\begin{equation}
	P(u,0) = \sum_{n=1}^{N}\sum_{m=1}^{M}w_{\substack{n,m}}e^{j\beta n d_x u}. \label{BP_u}
\end{equation}	
The grating lobes appeared along the $u$-axis at 

\begin{equation}
 	u_g = \pm n\lambda/d_x, n= 1,2, \cdots. \label{u_gb}
\end{equation}
Similarly, grating lobes appeared along the $v$-axis at    
\begin{equation}            
    v_g = \pm m\lambda/d_y, m= 1,2, \cdots. \label{v_gb}
\end{equation}
And grating lobes in the oblique direction were located at
\begin{equation}\label{45_gb}
    \begin{aligned}
	& u_g = \pm n\lambda/d_x, n= 1,2, \cdots,  \\
	& v_g = \pm m\lambda/d_y, m= 1,2, \cdots. 
\end{aligned}
\end{equation}
Since $P(0,0) = P(0+u_g, 0+v_g)$, the grating lobes were completely the same as the main lobe. Besides, whether these mainlobe-like grating lobes existed in possible visible region ($u,v \in [-2,2]$) merely depended on the elements pitches $d_x$ and $d_y$. Taking the first grating lobe as an example, it could be verified by the following

\begin{equation}\label{45_gb1}
	\begin{aligned}
		& |u_g| = |\pm \lambda/d_x| = \lambda/d_x \leq 2, if, d_x \geq 0.5\lambda.  \\
		& |v_g| = |\pm \lambda/d_y| = \lambda/d_y \leq 2, if, d_y \geq 0.5\lambda. 
	\end{aligned}
\end{equation}

In summary, if $d_x \geq 0.5\lambda$ or $d_y \geq 0.5\lambda$, mainlobe-like grating lobes would appear in the region ($u,v \in [-2,2]$). In this study, since $d_x = d_y = 0.58 \lambda$ and $|u_g|=|v_g|= 1.72$, five grating lobes were presented as shown in Fig. \ref{array_arrangement}. Thus, we could only seek trade-offs among main lobe width, SLL, and number of activated elements. 
If  $d_x \leq 0.5\lambda$ and $d_y \leq 0.5\lambda$, no grating lobes would exist in the region ($u,v \in [-2,2]$). However, small pitch usually corresponded to dividing an aperture into finer girds, which, in turn, posed limitations on the element size. It is worthwhile to perform further investigation in the future study of fine grid cases.

\subsection{Limitations and Future Work}

One limitation of the re-wiehgted L1 technique is that it is not guaranteed to reach a globally optimal solution for original problem \eqref{zeronorm} \cite{6193135}. Nevertheless, this techniques still dramatically increase the sparsity of the dense array (Fig. \ref{iteration_process}). Another limitation of formulating the sparse array synthesis problem as a convex optimization is the large memory consumption, especially for designing large sparse array cases. Here, large sparse array case refers to aperture with fine grids (e.g., $128\times 128$), not just aperture with large size, despite large apertures usually possess more elements. For 2D sparse array design, the number of optimization variable grows quadratically, which dramatically increases the computational complexity of solving SOCP by interior-point algorithm. Besides, the number of SOC constraints also influence the scale of the SOCP problem, which depends on the constrained area (e.g., \eqref{SLL_21dot2}) and discrete intervals, $\Delta u$ and $\Delta v$. The iterative optimization process in this work was conducted in Matlab 2019a on a PC workstation (Intel(R) Xeon(R) Gold 6136 CPU @3.00 GHz, 2.99GHz (2 processors), 192 GB RAM). 

Relatively low resolution, contrast and signal-to-noise ratio (SNR) are limitations to all the sparse arrays. Large aperture can be easily designed with the method adopted in this paper to improve the resolution. Non-linear beamforming methods, such as minimum variance (MV) \cite{4291510}, coherence factor \cite{li2003adaptive}, or deep-learning-based methods \cite{9138451, 10965879} can be adopted to improve the resolution and imaging contrast simultaneously. For the imaging SNR, compared with Spiral and Spiral-Taper, the Q-Flats produced more inferior SNR since apodization was applied on, which could be improved by coded excitation \cite{1406545, tamraoui2023complete}. 

Finally, the Q-Flats was designed by limiting the SLLs no higher than a certain level in this work. Other kinds of BP shape, for example, spiral-inspired BPs \cite{7549037}, could be also explored, which, however, increased the complexity of this study. This warrants further investigation.

\appendices

\section*{Acknowledgment}
We are very grateful to Prof.Piero Tortoli and Prof. Alessandro Ramalli for their great help in detemining the element distributions of spiral arrays on M3dV matrix array .

\bibliography{IEEEbibref}
\bibliographystyle{ieeetr}

\end{CJK}
\end{document}